\documentclass[aps,prd,twocolumn,superscriptaddress,amssymb,eqsecnum,showpacs,showkeyes,secnumarabic,graphics,floatfix,nofootinbib,tightenlines,longbibliography]{revtex4-1}
\usepackage{relsize}
\usepackage{blindtext}
\usepackage{enumitem}
\usepackage{xcolor}
\usepackage{graphicx}
\usepackage{epsfig}
\usepackage{bm}
\usepackage{cancel}
\usepackage{dcolumn}
\usepackage{amsmath}
\usepackage{hhline}
\usepackage[utf8]{inputenc}
\usepackage[breaklinks=true,colorlinks=true,
linkcolor=blue,urlcolor=blue,citecolor=cyan,
bookmarks=true,bookmarksopenlevel=2]{hyperref}

\usepackage{xcolor}
\usepackage[pass]{geometry}

\def \beq  {\begin{equation}}
\def \eeq  {\end{equation}}
\def \ber  {\begin{eqnarray}}
\def \eer  {\end{eqnarray}}

\begin{document}
\newcommand{\newc}{\newcommand}

\newc{\be}{\begin{equation}}
\newc{\ee}{\end{equation}}
\newc{\ba}{\begin{eqnarray}}
\newc{\ea}{\end{eqnarray}}
\newc{\bea}{\begin{eqnarray*}}
\newc{\eea}{\end{eqnarray*}}
\newc{\D}{\partial}
\newc{\ie}{{\it i.e.} }
\newc{\eg}{{\it e.g.} }
\newc{\etc}{{\it etc.} }
\newc{\etal}{{\it et al.}}
\newc{\lcdm}{$\Lambda$CDM }
\newcommand{\nn}{\nonumber}
\newc{\ra}{\Rightarrow}

\title{Existence and Stability of Static Spherical Fluid Shells in a Schwarzschild-Rindler-anti-de Sitter Metric}
\author{G. Alestas}\email{g.alestas@uoi.gr}
\author{G. V. Kraniotis}\email{gkraniotis@uoi.gr}
\affiliation{Department of Physics, University of Ioannina, 45110 Ioannina, Greece}

\author{L. Perivolaropoulos}\email{leandros@uoi.gr}
\affiliation{Department of Physics, University of Ioannina, 45110 Ioannina, Greece}

\date {\today}

\begin{abstract}
We demonstrate the existence of static stable spherical fluid shells in the Schwarzschild-Rindler-anti-de Sitter (SRAdS) spacetime where $ds^2 = f(r)dt^{2} -\frac{dr^{2}}{f(r)}-r^{2}(d\theta ^2 +\sin ^2 \theta d\phi ^2)$ with $f(r) = 1 -\frac{2Gm}{r} + 2 b r -\frac{\Lambda}{3}r^2$. This is an alternative to the well known gravastar geometry where the stability emerges due to the combination of the repulsive forces of the  interior de Sitter space with the attractive forces of the exterior Schwarzschild spacetime. In the SRAdS spacetime the repulsion that leads to stability of the shell comes from a negative Rindler term while the Schwarzschild and anti-de Sitter terms are attractive. We demonstrate the existence of such stable spherical shells for three shell fluid equations of state: vacuum shell ($p=-\sigma$), stiff matter shell ($p=\sigma$) and dust shell ($p=0$) where $p$ is the shell pressure and $\sigma$ is the shell surface density. We also identify the metric parameter conditions that need to be satisfied for shell stability in each case. The vacuum stable shell solution in the SRAdS spacetime is consistent with previous studies by two of the authors that demonstrated the existence of stable spherical scalar field domain walls in the SRAdS spacetime.

\end{abstract}
\maketitle

\section{Introduction}

Boundary layers (shells of matter sources with zero thickness) play an important role in both electromagnetism and general relativity. They provide a useful laboratory for the exploration of new phenomena while at the same time they approximate smooth solutions such as domain walls \citep{Vilenkin:1981zs, Kraus:1999it, PhysRevD.30.712, Perivolaropoulos:2018cgr, Alestas:2019wtw} or braneworlds \citep{Randall:1999ee, ArkaniHamed:1998rs, Antoniadis:1998ig, ArkaniHamed:1998nn, Langlois:2001uq}. Thin shells are also useful in describing gravitational collapse \citep{PhysRev.153.1388, Adler:2005vn, Dokuchaev:2010fv} or in constructing spherically symmetric vacuum solutions that avoid the presence of singularities (\eg gravastars \citep{Mazur:2004fk,Visser:2003ge,Lobo:2005uf,DeBenedictis:2005vp,Ansoldi:2008jw, doi:10.1142/S0218271820300049}).

Despite of the divergence of the stress-energy tensor on the thin shell, the corresponding singularities of the Einstein equations are mild and in fact they are easily integrable. Thus they lead to a simplification of the dynamical gravitational equations by converting the corresponding differential equations to finite difference equations known as 'junction conditions' \citep{Israel:1966rt,Blau:1986cw,Sen1924Jan}. These conditions lead to a determination of the discontinuities of various fields as the shell is crossed.

Thin spherical shells in General Relativity may be defined as 2+1 boundary hypersurfaces with energy momentum tensor $S^i_j\equiv\int^{R^+}_{R^-}T^i_j \; dr={\rm diag}(-\sigma,p,p)$, where $R$ is the shell radius, $r$ is the radial coordinate of the 3+1 dimensional spacetime,  $\sigma$ is the surface energy density and $p$ is the surface pressure on the shell hypersurface with equation of state $p=p(\sigma)$. The thin shell interpolates between an interior and an exterior spherically symmetric metric. The exterior metric is related to the interior metric in the context of the Israel junction conditions \cite{Israel:1966rt,Blau:1986cw,Sen1924Jan}. 

A well known spherical static stable  thin shell configuration corresponds to the gravastar that interopolates between an interior de Sitter metric and an exterior Schwarzschild metric and constitutes an extension of the Schwarzschild metric with eliminated singularity \citep{Mazur:2004fk,Visser:2003ge,Lobo:2005uf,DeBenedictis:2005vp,Ansoldi:2008jw}.

An alternative thin shell solution obtained using spherically symmetric scalar field  dynamical equations in a non-trivial background geometry has been obtained in Ref. \cite{Alestas:2019wtw}.  It was demonstrated that static metastable solutions can exist in the presence of a Schwarzschild-anti-deSitter curved spacetime \cite{Perivolaropoulos:2018cgr, Alestas:2019wtw} supplemented with the Rindler acceleration term. Thus the total metric is a Schwarzschild-Rindler-anti-deSitter (SRAdS) metric \cite{Grumiller:2010bz},
\begin{equation}
\begin{aligned}
ds^2 &= f(r)dt^{2} -\frac{dr^{2}}{f(r)}-r^{2}(d\theta ^2 +\sin ^2 \theta d\phi ^2)\\
f(r) &= 1 -\frac{2Gm}{r} + 2 b r -\frac{\Lambda}{3}r^2 
\end{aligned}
\label{Grummetric}
\end{equation}
where $b$ is the Rindler acceleration parameter and $\Lambda$ is the cosmological constant.

The metric (\ref{Grummetric}) has been constrained by solar system observations, indicating that $|b|<3nm/sec^2$ \cite{Carloni:2011ha, Iorio:2010tp} and it has been shown that it can lead to the production of flat rotation curves as well as contribute to the explanation \cite{Grumiller:2011gg, Iorio:2011zu} of the Pioneer anomaly \cite{Anderson:1998jd, Lammerzahl:2006ex} for $b > 0$. 

Such a metric including a linear term in $r$ is non-standard but it has been widely considered in the literature previously and  is physically motivated by at least three factors:
\begin{itemize}
\item In the context of GR a linear term in r it emerges naturally in a spherically symmetric metric,  in the context of a perfect fluid with density $\rho$ and pressure components  $\rho=-p_r=-2p_\theta =-2 p_\phi\sim 1/r$ \citep{Alestas:2019wtw}. In this sense it may be viewed as a generalization of the cosmological constant which gives a quadratic term in the metric and emerges in a spherically symmetric metric in the context of a homogeneous perfect fluid with  $\rho=-p_r=-p_\theta =- p_\phi= constant$. In fact any spherically symmetric metric given as a power series
\be 
f(r)=1-\sum^{N}_{n=-N}a_{n}r^{n}
\label{fpoly}
\ee
is supported by an energy - momentum tensor of the form
\begin{align}
T^{0}_{0} &=\frac{1}{\kappa}\sum^{N}_{n=-N}a_{n}(1+n)r^{n-2}=\rho \label{rho}\\
T^{r}_{r} &=T^{0}_{0}=-p_{r} \label{p_r}\\
T^{\theta}_{\theta} &=\frac{1}{2\kappa}\sum^{N}_{n=-N}a_{n}n(1+n)r^{n-2}=-p_{\theta} \label{ptheta}\\
T^{\phi}_{\phi} &=T^{\theta}_{\theta}=-p_{\phi} \label{pPhi} 
\end{align}
For a linear term in the metric ($n = 1$), the corresponding energy-momentum term varies as $1/r$ which leads to asymptotic flatness since the energy-momentum tensor vanishes at infinity. For  $n = 2$ we have constant energy density-pressure via the cosmological constant term and for the case of $n = -1$ we have the vacuum solution which corresponds to zero energy momentum tensor.
\item It is the spherically symmetric vacuum solution in various modified gravity theories including Weyl \citep{Mannheim:1988dj} and in 2 dimensional scalar-tensor theories \citep{Grumiller:2011gg, Perivolaropoulos:2019vgl}. In these theories, the proper sign of the linear term can lead to additional attractive gravity that can play the role of dark matter without actual existence of any form of energy-momentum \cite{Carloni:2011ha, Iorio:2010tp, Mannheim:1988dj}.
\item  In view of the generic and natural existence of the  terms proportional to  $1/r$  (GR vacuum) and $r^2$  (cosmological constant) the presence of a linear term $\sim r$ emerges as a natural generalization with potentially interesting physical effects.  One of these effects is the existence of scalar hair (stable spherical scalar domain wall (demonstrated in Ref. \cite{ Perivolaropoulos:2018cgr}).
\end{itemize}

The metric's property of supporting metastable spherical domain walls motivates the search of additional stable shell solutions described as general fluid thin shells as opposed to scalar field vacuum energy shells (domain walls).  Such an analysis would be based on the Israel junction conditions formalism as opposed to the solution of dynamical scalar field equations. The following questions therefore emerge:
\begin{itemize}
\item
Are there static, stable fluid shell solutions in a SRAdS background geometry?
\item
If yes, what are the conditions for their stability given the equation of state of the fluid shell?
\item
What is the metric parameter range for shell stability and how does the stability radius change as a function of these parameters?
\end{itemize}

These questions will be addressed in the present analysis. We implement the Israel junction conditions in the context of a fixed equation of state of the fluid shell and a SRAdS background metric with a discontinuous value of  $m$ across the shell and  fixed values of $b$ and $\Lambda$ with no discontinuity as the shell is crossed. We thus derive the stability conditions and identify the range of metric parameters $b,\Lambda$, that satisfy these conditions for given values of the shell coordinate radius $R$, shell surface density $\sigma$ and  mass parameters inside and outside the shell ($m_{-}$, $m_{+}$). The conditions that need to be satisfied for stability by the shell density and shell radius are also determined.

The structure of this paper is the following: In the next section we develop the general formalism for the derivation of stability conditions by implementing the Israel junction conditions on the SRAdS metric for a shell with a general fluid equation of state. In section \ref{sectionIII} we consider three specific applications of the method for corresponding shell fluid equations of state: vacuum shell, stiff matter shell and matter shell and find the particular stability conditions and parameter regions in each case. Finally in section \ref{sectionIV} we conclude summarize and discuss possible extensions of this analysis. 

In what follows we set $G=c=1$. In most cases we will also set the interior mass parameter $m_- = 1$. Thus in this context, a dimensionless form of $\Lambda$ corresponds to the dimensionless combination $m_-^2 \Lambda$. Furthermore, whenever Greek letters are used as indices they correspond to spacetime ones, while Roman (Latin) indices range over the coordinates of the 2+1-surface of the shell. Also, the radius of the shell is always considered in the region outside the event horizon of the black hole. Notice that for the parameter values considered (AdS) there is no cosmological horizon but only an event horizon.

\section{Thin Shells: Existence and Stability}\label{sectionII}
Consider a thin spherical shell with coordinate radius R interpolating between an interior ($g_{\mu\nu}^-$) and an exterior metric ($g_{\mu\nu}^+$). Let the interior and exterior metrics be of the form \cite{Visser:2003ge,Mazur:2004fk,Frauendiener_1990},
\begin{widetext}
\ba
ds^2 = f_{\pm}(r_{\pm})dt^{2} -\frac{dr_{\pm}^{2}}{f_{\pm}(r_{\pm})}-r_{\pm}^{2}(d\theta ^2 +\sin ^2 \theta d\phi ^2)
\ea
\end{widetext}
where,
\ba
f_{\pm}(r_{\pm}) = 1 - \frac{2m_{\pm}(r_{\pm})}{r_{\pm}} \label{fr}
\ea
and
\ba
m_{\pm}(r_{\pm}) = m_{\pm}-br_{\pm}^2 +\frac{\Lambda}{6}r_{\pm}^3 .\label{m_pm}
\ea

We now impose the following conditions:
\begin{enumerate}
\item
Continuity of the metric on the shell $(r_{-} = r_{+} = R)$. This implies 
\ba
f_{+}(r_{+})dt_{+}^{2} -\frac{dr_{+}^{2}}{f_{+}(r_{+})} = f_{-}(r_{-})dt_{-}^{2} -\frac{dr_{-}^{2}}{f_{-}(r_{-})}. \label{cont_eq}
\ea
which leads to
\begin{align}
t_{-} &= \frac{f_{+}(R)}{f_{-}(R)}t_{+}\\
\frac{dr_{-}}{dr_{+}} &= \frac{f_{-}(R)}{f_{+}(R)}.
\end{align}
\item
The Israel junction conditions \cite{Israel:1966rt} expressed through a discontinuity of the extrinsic curvature on the shell hypersurface $\Sigma$. The extrinsic curvature (second fundamental form) at either side of the three-dimensional (2+1) hypersurface $\Sigma$ swept by a spherically symmetric shell, embedded in the four-dimensional spacetime are:
\ba
K_{ij}^{\pm}=\left(h_{\mu}^{\lambda}n_{\nu;\lambda}\frac{dx^{\mu}}{dx^i}\frac{dx^{\nu}}{dx^j}\right)^{\pm}_{\Sigma},
\ea
where $x^i$ are coordinates on $\Sigma$, $h_{\mu\nu}=g_{\mu\nu}-n_{\mu}n_{\nu}$, and $(;)$ denote covariant derivative with respect to $g_{\mu\nu}^{\pm}$. Let
\ba
g(x^{\alpha}(x^i))=0,
\ea 
denote the parametric equation for $\Sigma$ as embedded in the 4-dimensional spacetime. The unit 4-normals to $\Sigma$ in the four-dimensional spacetime are given by \cite{Sen1924Jan},
\ba
n_{\alpha}=\pm\left(\left|g^{\beta\gamma}\frac{\partial g}{\partial x^{\beta}}\frac{\partial g}{\partial x^{\gamma}}\right|\right)^{-1/2}\frac{\partial g}{\partial x^{\alpha}},
\ea
We assume $n_{\alpha}\not =0$ and label $\Sigma$ as timelike for $n_{\alpha}n^{\alpha}=1$ (a spacelike normal). The Israel junction conditions are expressed as discontinuities of the extrinsic curvature of the shell
\be 
\left[\left[ K_{ij}\right]\right]=-8\pi \left[S_{ij} - \frac{1}{2} S h_{ij}\right ]
\label{israelc1}
\ee
where $\left[\left[X\right]\right]\equiv \left[X_+\right]-\left[X_-\right]$ denotes the discontinuity of the quantity $X$ as the shell is crossed.

In the case of a static shell in the SRAdS metric (\ref{Grummetric}) the extrinsic curvature tensor takes the form
\ba
K_{ij} = \sqrt{f_{\pm}(r_{\pm})}{\rm diag}\left(\frac{\frac{1}{2}f_{\pm}^{\prime}(r_{\pm})}{f_{\pm}(r_{\pm})},\frac{1}{r_{\pm}},\frac{1}{r_{\pm}}\right).\label{curvature}
\ea
and the Israel junction conditions for a dynamic shell are of the form \cite{Visser:2003ge}
\begin{align}
\sigma &= -\frac{1}{4\pi R}\left[\left[\sqrt{1-2m_{\pm}(R)/R +\dot{R}^2}\right]\right]\label{surfdens}\\
p &= \frac{1}{8\pi R}\left[\left[\frac{1-m_{\pm}(R)/R -m_{\pm}(R)' +\dot{R}^2 +R\ddot{R}}{\sqrt{1-2m_{\pm}(R)/R +\dot{R}^2}}\right]\right]
\end{align}
where ($^{\prime}$) denotes derivative of $m_{\pm}(r)$ with respect to $r$ at $r=R$ and dot denotes  derivative with respect to the proper time of the shell defined as

\begin{widetext}
\ba
d\tau^2 &=&
\;\left[1-{2m_{\pm}(R)\over R}\right]\; {dt}^2
-\;{1\over1-2m_{\pm}(R)/R} \left[{dR\over dt}\right]^2 dt^2
\ea
\end{widetext}
\end{enumerate}

These equations lead also to the energy conservation equation on the shell,
\ba
\frac{d}{d\tau}(\sigma R^2) + p\frac{d}{d\tau}R^2 = 0 \label{energcons}
\ea

The eq. (\ref{surfdens}) may also be expressed as,
\ba
\frac{1}{2}\dot{R}^2 + V(R) = E \label{energconspart}
\ea
where,
\begin{widetext}
\ba
V(R) \equiv 1+\frac{4\;m_{+}(R)\;m_{-}(R)}{16\pi^2 \sigma^2 R^4}-\left[\frac{4\pi \sigma R^2}{2R}+\frac{m_{+}(R)+m_{-}(R)}{4\pi \sigma R^2}\right]^2 \label{Master}
\ea
\end{widetext}
and $E=0$.

Clearly,  eq. (\ref{energconspart}) is identical with the energy  conservation equation of a particle moving in one dimension with coordinate $R(\tau)$ and zero energy. Thus, the conditions for the existence of a static, stable shell may be written as,

\begin{align}
\begin{split}
V(R)=0 \\
V^{'}(R)=0 \\
V^{''}(R)>0
\end{split}
\label{system_cond}
\end{align}

These conditions, along with the equation of state $p(\sigma)$ and the energy conservation eq. (\ref{energcons})  may be used to identify constraints on the metric parameters appearing in the expressions of $m_{-}(R)$ and $m_{+}(R)$ required for the existence of a stable spherical shell with given radius R. In the present analysis we consider the particular forms of $m_\pm(r)$  given by eq. (\ref{m_pm}) corresponding to the SRAdS metric.  In this case the potential of eq. (\ref{Master}) takes the form,

\begin{widetext}
\ba
V(R) = 1 - \frac{m_{-}+m_{+}}{R}+2bR-\frac{\Lambda R^2}{3}-\frac{(m_{-}-m_{+})^2}{16\pi^2 R^4 \sigma(R)^2}-4\pi^2\sigma(R)^2 R^2
\label{potsrads}
\ea
\end{widetext}

In the context of a constant shell fluid equation of state we have $p=w\sigma$ and it is easy to show that energy conservation (\ref{energcons}) leads to 
\be 
\sigma = \sigma_0' \;\left( \frac{R}{R_0}\right)^{-2(w +1)}
\label{sigmawr}
\ee
where $\sigma_0'$ is the surface density of a shell of radius $R_0$. In what follows we define 
\be
\sigma_0\equiv \sigma_0' R_0^{2(w+1)}.\label{eq:sigma0dimension}
\ee

For example in the special case of a pressureless matter shell ($w=0$) we obtain the expected result $\sigma(R)\sim R^{-2}$ while for a vacuum shell we have $\sigma(R)=\sigma_0=const$. The dimensionality of $\sigma_{0}$ is therefore depended on the equation of state parameter $w$, resulting in different dimensions for $\sigma_0$ regarding each case of $w$ discussed in the next section.
 
In the special case when $\sigma$ is independent of R discussed in the next section (vacuum shell) it is straightforward to show that a minimum of the potential (\ref{potsrads})  exists for $b<0$ and $\Lambda<0$ due to the attractive nature of the linear potential term $2\; b\;R$ which dominates at large $R$ competing with the repulsive effects of the quadratic potential term $-\Lambda R^2/3$ which dominates at even larger $R$. 

For more general metrics or fluid equations of state than the one considered here it is clearly possible to have several minima for the potential corresponding to configurations of more than one stable concentric shells. 

In the next section we identify the metric parameter ranges of $b$ and $\Lambda$ that allow for stable shells in the spacial cases of three shell fluid equations of state.
\begin{figure*}[h!t]
\centering
\begin{center}
$\begin{array}{@{\hspace{-0.10in}}c@{\hspace{0.0in}}c}
\multicolumn{1}{l}{\mbox{}} &
\multicolumn{1}{l}{\mbox{}} \\ [-0.2in]
\includegraphics[scale=0.65]{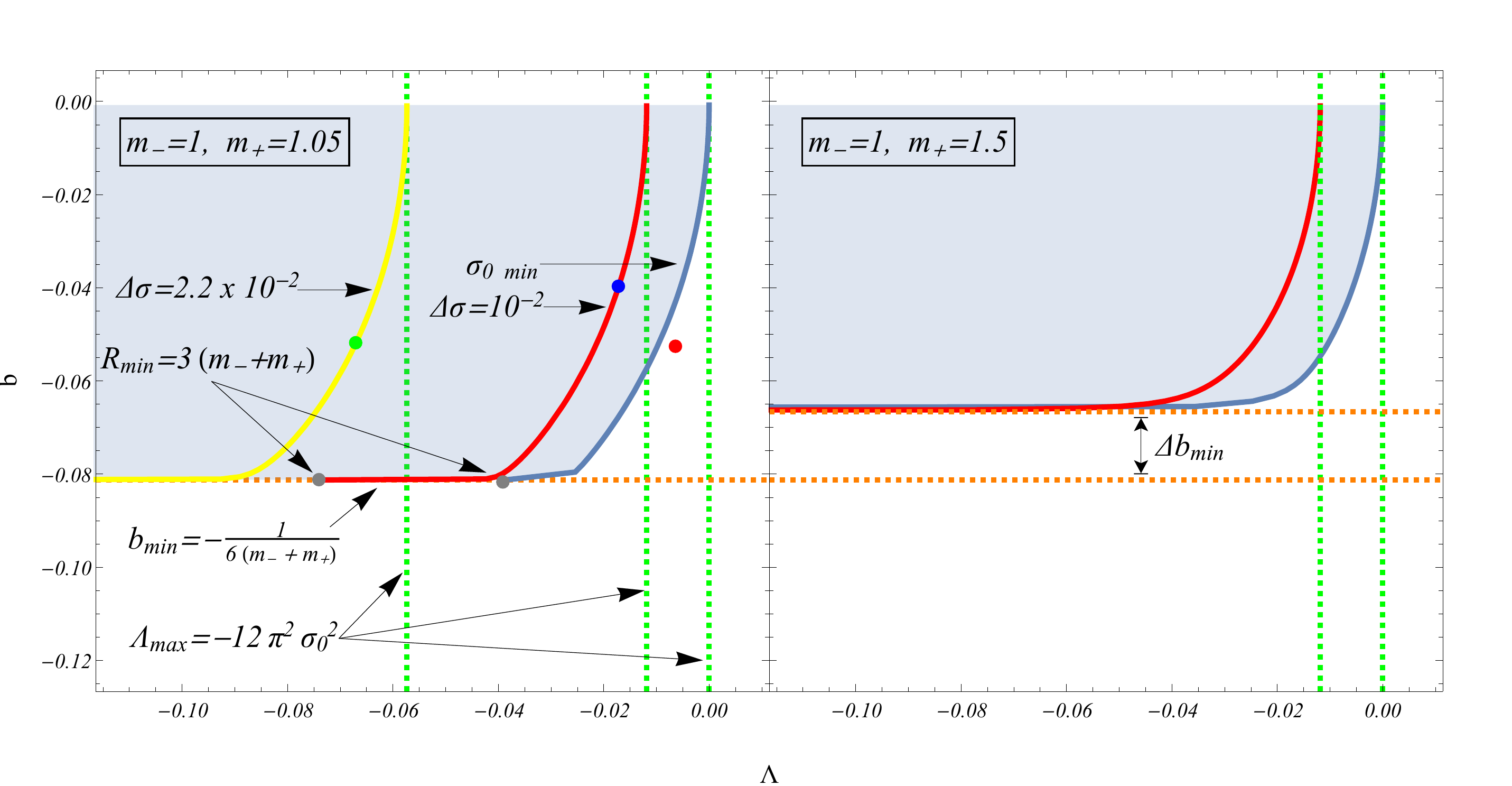} \\
\end{array}$
\end{center}
\vspace{0.0cm}
\caption{\small The shell stability region (light blue region) in the $b-\Lambda$ parameter space for two pairs of $m_+-m_-$. The colored curves correspond to fixed value of surface density in the stability range $\sigma_0\equiv \sigma_{0min}+\Delta \sigma >\sigma_{0min}$ while $R$ varies such that $R>R_{min}$. Since we have set $m_-=1$, the cosmological constant becomes dimensionless and equal to the product $m_-^2 \Lambda$. We thus study two separate cases with different values for the exterior mass $m_+$. In the right panel we set $m_+/m_- = 1.5$ and in the left panel at $m_+/m_- = 1.05$. We can clearly see that compared to the $m_+/m_-  = 1.05$ one, the shell which corresponds to the $m_+/m_-  = 1.5$ exterior mass displays a greater value of the lower boundary of the b parameter ($b_{min}$) as well as a smaller value of the $R_{min}$ limit.}
\label{fig1}
\end{figure*}
We then proceed by finding the ranges of the parameters $b, \Lambda$ which allow for stable spherical shell solutions, via the implementation of these conditions for different cases of interior and exterior equations of state.
\section{Special Cases}
\label{sectionIII}

\subsection{Vacuum fluid shell ($w=-1$)}

The simplest case of a stable spherical shell is obtained assuming a vacuum fluid equation of state
\ba
p = -\sigma.
\ea
This case is similar to the case of a stable domain wall in the SRAdS metric discussed in  \citep{Alestas:2019wtw} using theoretical methods. It was shown that such metastable topological field configurations may indeed exist for $b<0$, $\Lambda<0$ due to the competing attractive-repulsive effects of the linear and quadratic terms of the metric functions. In the vacuum fluid case we have from eq. (\ref{sigmawr})
\ba
\sigma(R) = \sigma_{0} = const.
\ea

where the $\sigma_0$ has dimensions of $R^{-1}$, in accordance with eq. (\ref{eq:sigma0dimension}). In this case the system (\ref{system_cond}) becomes,

\begin{widetext}
\ba
&&V(R) = 1 - \frac{m_{-}+m_{+}}{R}+2bR-\frac{\Lambda R^2}{3}-\frac{(m_{-}-m_{+})^2}{16\pi^2 R^4 \sigma_{0}^2}-4\pi^2\sigma_{0}^2 R^2 = 0
\label{vacV}\\
&&\frac{\partial V}{\partial r}\Bigl|_{r=R} = 2b +\frac{m_{-}+m_{+}}{R^2}-\frac{2\Lambda R}{3}
+\frac{(m_{-}-m_{+})^2}{4\pi^2 R^5 \sigma_{0}^2}-8\pi^2\sigma_0^2 R = 0
\label{der1vacV}\\
&&\frac{\partial^2 V}{\partial r^2}\Bigl|_{r=R} = - \frac{2\Lambda}{3} - \frac{2(m_{-}+m_{+})}{R^3}-\frac{5(m_{-}-m_{+})^2}{4\pi^2 R^6 \sigma_0^2} - 8\pi^2\sigma_0^2 > 0.
\label{der2vacV}
\ea
\end{widetext}

The solution of the system (\ref{vacV} - \ref{der2vacV})may be written as
\begin{widetext}
\ba
\Lambda(R,\sigma_0) &&= \frac{15 (m_{-}-m_{+})^2}{16 \pi^2 R^6 \sigma_0^2}+\frac{6 (m_{-}+m_{+})-3
   R}{R^3}-12 \pi ^2 \sigma_0^2 \label{lr}\\
b(R,\sigma_0) &&= \frac{3(m_{-}-m_{+})^2 + 8\pi^2 [3 (m_{-}+m_{+})-2R]R^3 \sigma_0^2}{16\pi^2 \sigma_0^2 R^5}\label{br}\\
R &&> 3(m_{-} + m_{+}) \equiv R_{min}\label{R_min} \\
\sigma_0 &&\equiv \frac{\sqrt{15} \sqrt{-\frac{(m_{-}-m_{+})^2}{R^3 (3 m_{-}+3 m_{+}-R)}}}{4\pi} + \Delta\sigma > \frac{\sqrt{15} \sqrt{-\frac{(m_{-}-m_{+})^2}{R^3 (3 m_{-}+3 m_{+}-R)}}}{4\pi } \equiv \sigma_{0min}\label{sigma_min}
\ea
\end{widetext}
where $\Delta\sigma$ allows for small perturbations on the surface density, higher than that of its minimum value $\sigma_{0min}$.

\begin{figure}[h!]
\centering
\includegraphics[scale=0.3]{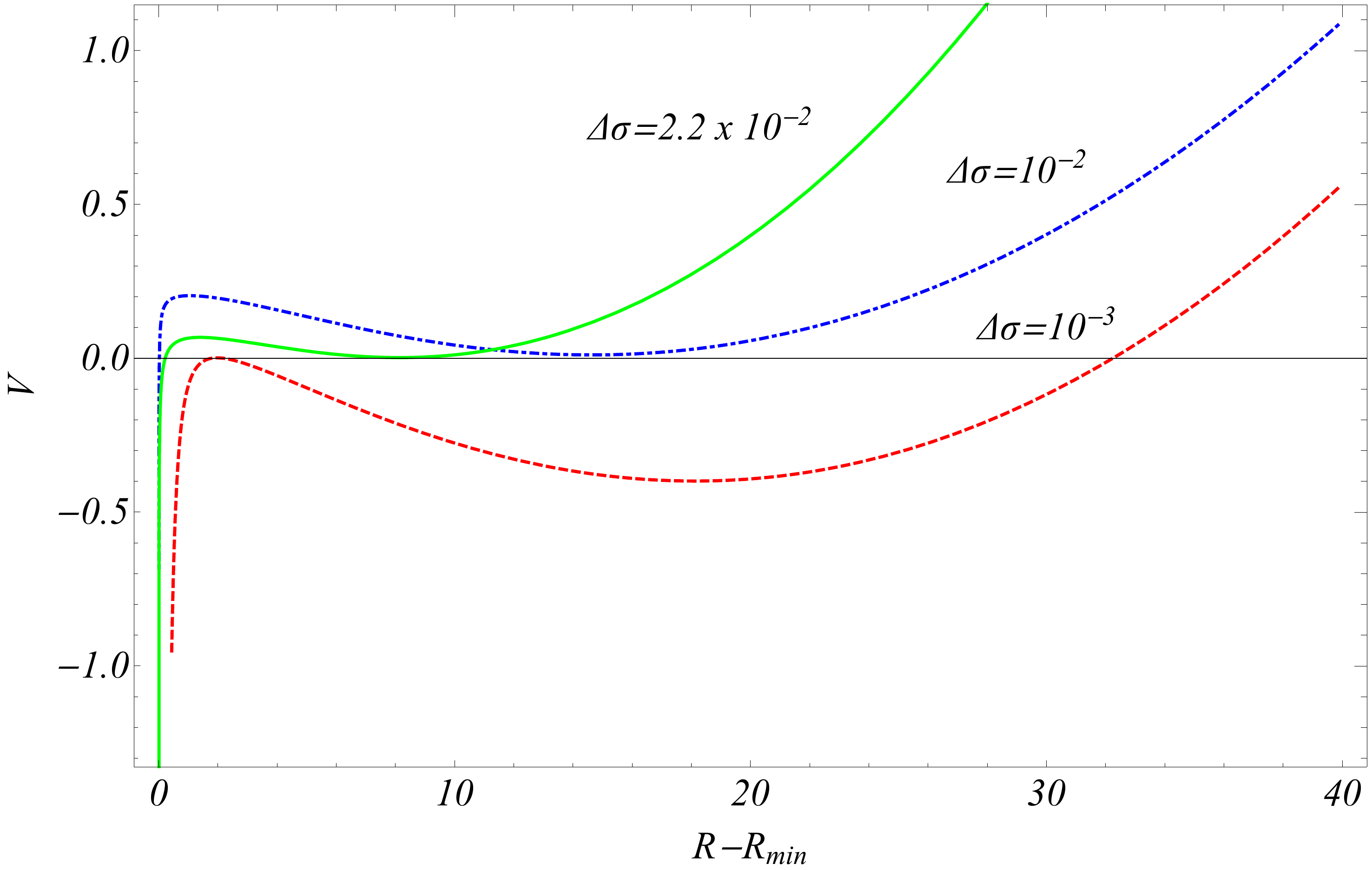} \\
\caption{\small The potential (\ref{vacV}) for parameter values corresponding to the three points shown in Fig. \ref{fig1}. These points correspond to parameter values: $(R=14.28, \Delta\sigma=2.2\times10^{-2},  b=-5.18\times10^{-2},  \Lambda=-6.76\times10^{-2})$ (green point), $(R=20.78, \Delta\sigma=10^{-2},  b=-3.97\times10^{-2},  \Lambda=-1.72\times10^{-2})$ (blue point) and $(R=7.95, \Delta\sigma=10^{-3}, b=-5.25\times10^{-2}, \Lambda=-6.39\times10^{-3})$ (red point). Notice that the red point which is outside the stability region corresponds to a potential which does have an extremum with $V(R)=0$ (for $R\simeq 2$) which implies the existence of a shell solution. However, this extremum corresponds to a local maximum indicating instability of the corresponding shell solution.}
\label{fig2}
\end{figure}

\begin{figure}[h!]
\centering
\includegraphics[scale=0.3]{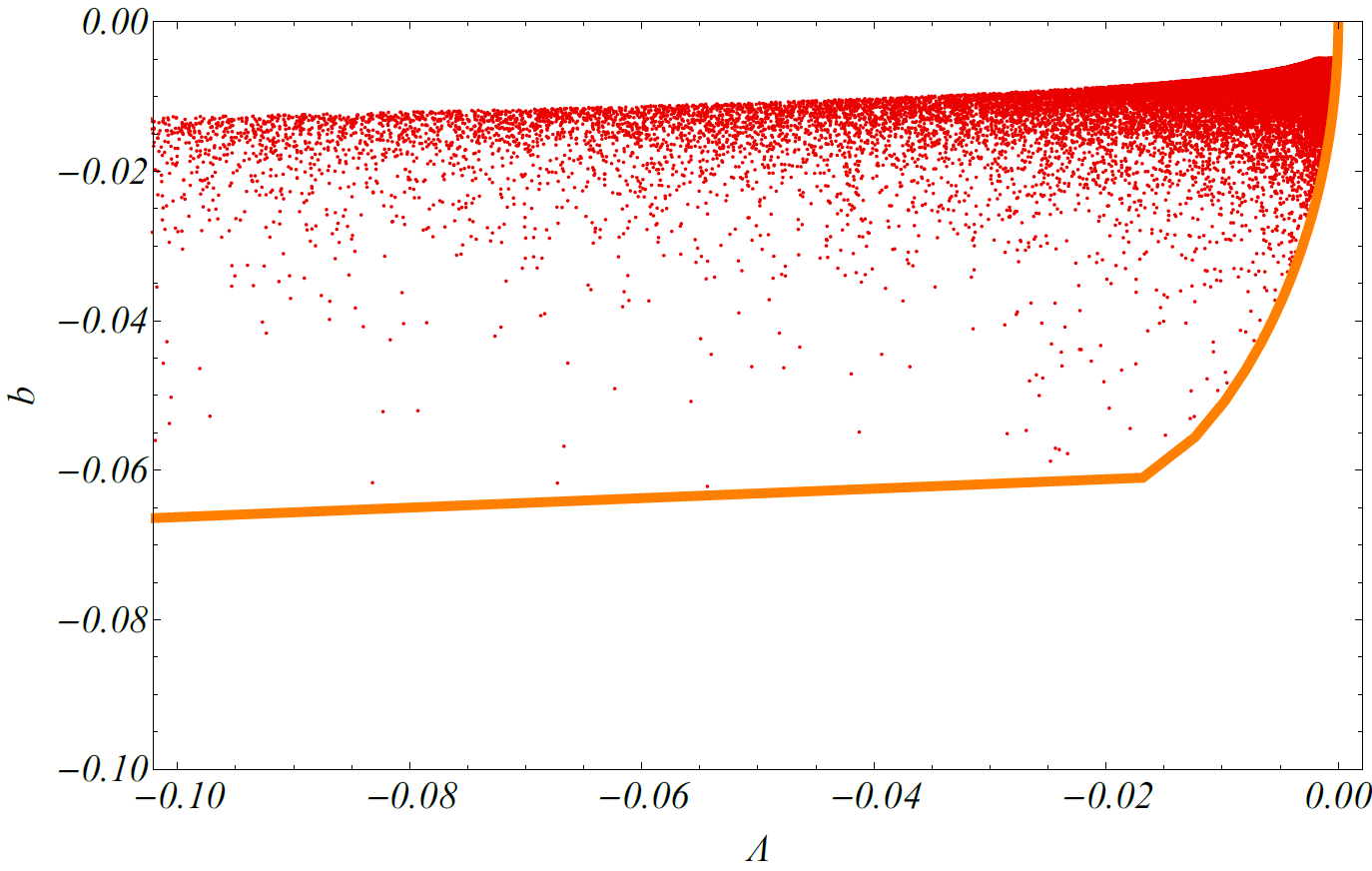} \\
\caption{\small A random Monte-Carlo selection of points that satisfy the shell existence and stability conditions (\ref{lr} - \ref{sigma_min}) for $m_{+}=1.5$. The orange line represents the limit of the region which is clearly respected by all the randomly selected points which span the stability region.}
\label{fig3}
\end{figure}

The existence of lower limits on the values of R and $\sigma_0$ allows the analytical derivation of the boundaries in the $b, \Lambda$ parameter space of the region that permits a stable shell solution. In particular when the shell radius takes its lower limit value $R=R_{min}$, we have $\sigma_{0min} = \infty$ and $\Lambda \rightarrow -\infty$  which implies the existence of a low bound on b for large $|\Lambda |$ as
\ba
\Lambda \rightarrow -\infty \implies b \rightarrow -\frac{1}{6(m_{+}+m_{-})}.\label{b_bound}
\ea
Similarly for large shell radius $(R \rightarrow \infty)$ we have,
\ba
\sigma_{0min} \rightarrow 0 \text{ and } \Lambda \rightarrow -12\pi^2 \sigma_{0}^2 > 0.
\ea
From eq. (\ref{br}) implies the existence of \ an upper bound for the parameter b,
\ba
b < 0 \text{ with } \Lambda < \Lambda_{max} = -12\pi^2 \sigma_{0}^2.\label{l_bound}
\ea

These analytically derived boundaries of the stability parameter region may be displayed by showing contours in the $(b,\Lambda)$  parameter space that show the shell stability regions in the context of the constraints (\ref{lr} - \ref{sigma_min}) for fixed values of $m_+, m_-$. Clearly, the boundaries expressed by eqs (\ref{b_bound} - \ref{l_bound}) are respected by these regions as demonstrated in Fig. \ref{fig1}. As expected (right panel of Fig. \ref{fig1}), the minimum value of $b$ in the stability region increases as $m_+$ is increased (see eq. (\ref{b_bound})). In Fig. \ref{fig2} we show the form of the potential (\ref{Master}) for three sets of parameters $(R,\sigma,b,\Lambda)$ inside and outside the stability region of Fig. \ref{fig1}. As expected the potential develops a minimum with $V(R) = 0$ only for the parameters inside the stability region while the parameter values in the instability region correspond only to a local maximum of the potential at the corresponding value of R.

In order to illustrate the validity of the stability boundaries shown in Fig. \ref{fig1} we show a random set of stability parameter points in Fig. \ref{fig3} which is constructed as follows:
\begin{enumerate}
\item
We fix $m_{-}=1$, $m_{+}\equiv m_+/m_-=1.5$. ($m_-$ and $m_+$ become dimensionless since we have set $m_-=1$).  Then we construct the stability boundary as the set of points with $b=b(R,\sigma_{0min}(R))$, $\Lambda=\Lambda(R,\sigma_{0min}(R))$, where $R>R_{min}$ (see eq.(\ref{R_min})), $\sigma_{0min}(R)$ is obtained from eq. (\ref{sigma_min}).
\item
We construct  a random selection of shell radius values $R_i$ respecting the stability constraint (\ref{R_min}). For each value of $R=R_i$ we consider a random value for $\sigma_i$ such that $\sigma_i>\sigma_{0min}(R_i)$ (see eq. (\ref{sigma_min})). For the given random pair $(R_i,\sigma_i)$ we obtain the stability parameters $(\Lambda,b)$ and plot the corresponding point in Fig. \ref{fig3}.
\item
We repeat this process for $i=1,...,N$ $(N=5\times10^{4})$ thus constructing Fig. \ref{fig3}.
\end{enumerate}
Clearly all the points corresponding to stable shell parameter values are within the stable region thus testing the validity of this region and the consistency of Fig. \ref{fig1}.

\subsection{Stiff matter fluid shell ($w=1$)}
\label{secstiff}

A stiff matter shell has equation of state
\be
p=\sigma.
\ee
From the eq. (\ref{sigmawr}) with $w=1$ we obtain
\be
\sigma(R)=\sigma_0 R^{-4}  \label{eq_stiff1}
\ee
where $\sigma_0$ has dimensions of $R^3$, in accordance with eq. (\ref{eq:sigma0dimension}). For this equation of state the potential (\ref{potsrads})  takes the form,

\begin{widetext}
\ba
V(R) = 1+ 2b R-\frac{\Lambda R^2}{3}-\frac{(m_{-}-m_{+})^2 R^4}{16\pi^2 \sigma_{0}^2}-\frac{m_{-}+m_{+}}{R}-\frac{4\pi^2 \sigma_{0}^2}{R^6}\label{eq_stiff}
\ea
\end{widetext}

The system of stability conditions (\ref{system_cond}) in this case takes the form,
\begin{widetext}
\ba
&&V(R) = 1 - \frac{m_{-}+m_{+}}{R}+2bR-\frac{\Lambda R^2}{3}-\frac{(m_{-}-m_{+})^2 R^4}{16\pi^2\sigma_{0}^2}-\frac{4\pi^2\sigma_{0}^2}{R^6}= 0
\label{firderV}\\
&&\frac{\partial V}{\partial r}\Bigl|_{r=R} = 2b +\frac{m_{-}+m_{+}}{R^2}-\frac{2\Lambda R}{3}
-\frac{(m_{-}-m_{+})^2 R^3}{4\pi^2\sigma_{0}^2}+\frac{24\pi^2\sigma_0^2}{R^7} = 0
\label{secderV}\\
&&\frac{\partial^2 V}{\partial r^2}\Bigl|_{r=R} = - \frac{2\Lambda}{3} - \frac{2(m_{-}+m_{+})}{R^3}-\frac{3(m_{-}-m_{+})^2 R^2}{4\pi^2\sigma_0^2}-\frac{168\pi^2\sigma_0^2}{R^8}>0
\label{2derV}
\ea
with solution for existence of shell solution
\ba
\Lambda(R,\sigma_{0}) &&= -\frac{9(m_{-}-m_{+})^2 R^2}{16\pi^2 \sigma_{0}^2}+\frac{6(m_{+}+m_{-})-3R}{R^3}+\frac{84\pi^2 \sigma_{0}^2}{R^8}\label{lr_stiff}\\
b(R,\sigma_{0}) &&= -\frac{(m_{-}-m_{+})^2 R^3}{16\pi^2 \sigma_{0}^2}+\frac{3(m_{-}+m_{+})-2R}{2R^2}+\frac{16\pi^2 \sigma_{0}^2}{R^7}\label{br_stiff}
\ea
\end{widetext}
The stability condition (\ref{2derV}) leads to the constraints
\begin{widetext}
\ba
- 4\pi\sqrt{\sigma_0^2(R^6-6m_{+}R^5-100\pi^2\sigma_0^2)} &&<\frac{3[(m_{-}-m_{+})R^5+8\pi^2\sigma_0^2]}{\sqrt{3}} <4\pi\sqrt{\sigma_0^2(R^6-6m_{+} R^5-100\pi^2\sigma_0^2)},\label{stiffcond1} \\
R^6 &&>6m_{+} R^5+100\pi^2\sigma_{0}^{2}.\label{stiffcond2}
\ea
\end{widetext}
which must be met simultaneously in order for a stability region to exist.

Using again the Monte-Carlo method of Fig. \ref{fig3} with random values of $R$ and $\sigma_0$ in the region allowed by eqs. (\ref{stiffcond1})-(\ref{stiffcond2}), we obtain the corresponding stability values of $\Lambda$ and $b$ which map the stability region shown in Fig. \ref{fig4}.

The range of the $(b,\Lambda)$ parameters for which we have stable solutions for the stiff matter case appears to be significantly narrower than the corresponding one for the case of vacuum shell. The reduction of the stability region in this case is due to the repulsive term of the potential of eq. (\ref{eq_stiff}) proportional to $R^4$  which is not present in the vacuum shell case and  spoils the attractive effects of the anti-deSitter term $\sim \Lambda R^2$  ($\Lambda<0$)  needed for the formation of a potential minimum at large R.

\subsection{Pressureless dust fluid shell ($w=0$)}

For a pressureless dust fluid shell we have $p=0$ and eq. (\ref{energcons}) leads to a surface energy density of the form
\be
\sigma(R)= \sigma_0 R^{-2}
\ee

with $\sigma_0$ dimensions of $R$, in accordance with eq. (\ref{eq:sigma0dimension}). In this case the potential takes the form,

\begin{widetext}
\ba
V(r) =1+ 2b R-\frac{\Lambda R^2}{3}-\frac{(m_{-}-m_{+})^2}{16 \pi ^2
   \sigma_{0} ^2}-\frac{m_{-}+m_{+}}{R}-\frac{4 \pi ^2 \sigma_{0} ^2}{R^2}\label{eq_ordin}
\ea
Solving the system (\ref{system_cond}) for this potential yields the following forms for $\Lambda$ and $b$ (existence conditions)
\ba
\Lambda(R,\sigma_{0}) &&= \frac{3 (m_{-}-m_{+})^2}{16 \pi ^2 R^2 \sigma_{0} ^2}+\frac{6 (m_{-}+m_{+})-3R}{R^3}+\frac{36 \pi ^2 \sigma_{0} ^2}{R^4}\label{lr_ordin}\\
b(R,\sigma_{0}) &&= \frac{(m_{-}-m_{+})^2}{16 \pi ^2 R \sigma_{0} ^2}+\frac{3 (m_{-}+m_{+})-2R}{2R^2}+\frac{8 \pi ^2 \sigma_{0} ^2}{R^3}\label{br_ordin}
\ea
\end{widetext}

Since the dimensionality of $\sigma_{0}$ is $R$ it is evident that eqs. \eqref{eq_ordin} - \eqref{br_ordin} are dimensionally correct ($\Lambda \sim R^{-2}$, $b \sim R^{-1}$). While stability of the shell implies that

\begin{widetext}
\ba
\|24\pi^2\sigma_0^2+(m_{-}-m_{+})R\| &&< 4\pi\sqrt{\sigma_0^2(12\pi^2\sigma_0^2-6m_{+}R+R^2)},\\
0 &&<12\pi^2\sigma_0^2-6m_{+} R+R^2.
\ea
\end{widetext}

\begin{figure}[h!]
\centering
\includegraphics[scale=0.3]{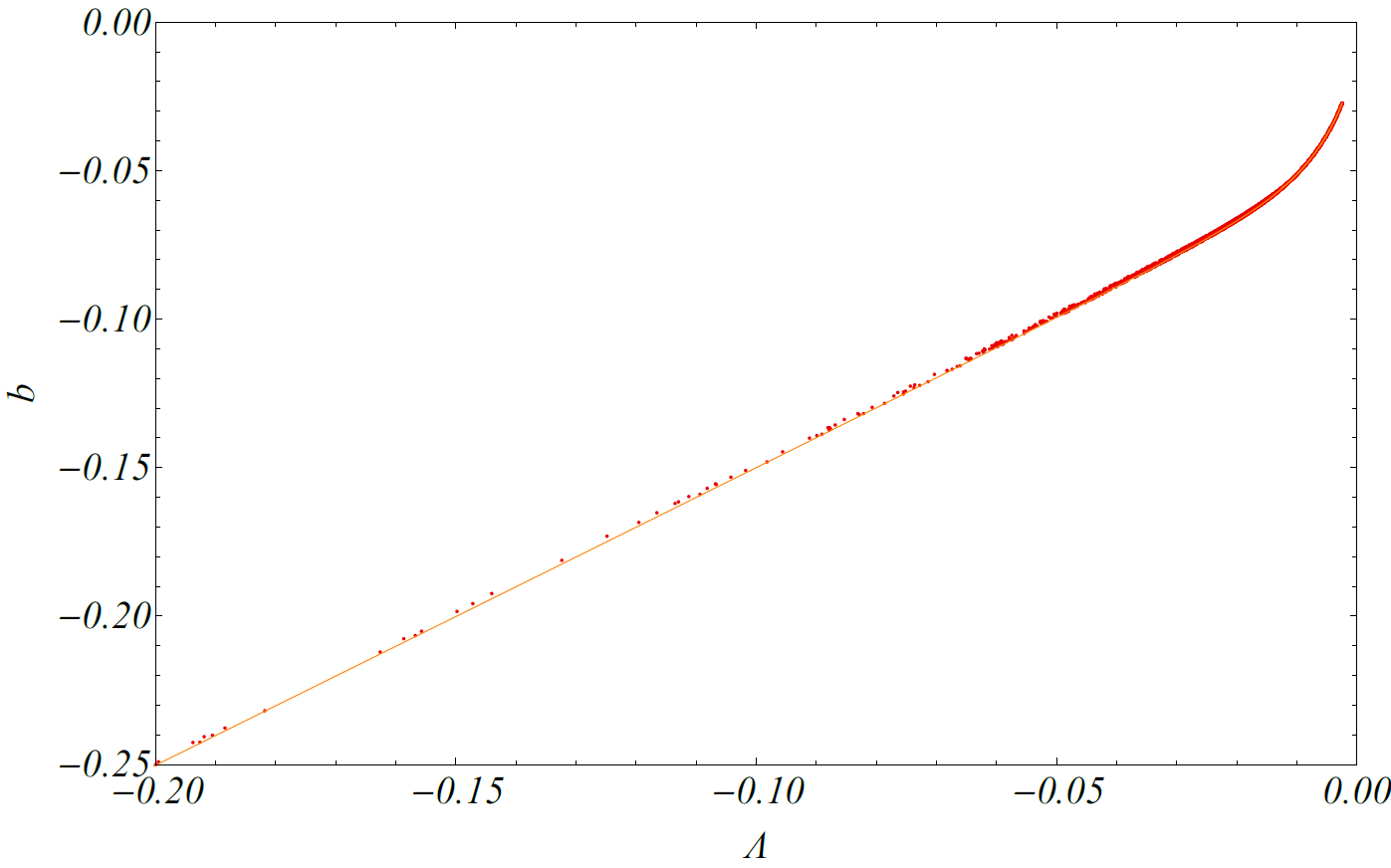} \\
\caption{\small A Monte-Carlo map of the stiff matter shell stability parameter region $(b,\Lambda)$ for  $m_{-}=1$, $m_{+}=1.5$. Notice that in this case the stability parameter range is much more narrow than in the case of the vacuum shell. }
\label{fig4}
\end{figure}

\begin{figure}[h!]
\centering
\includegraphics[scale=0.3]{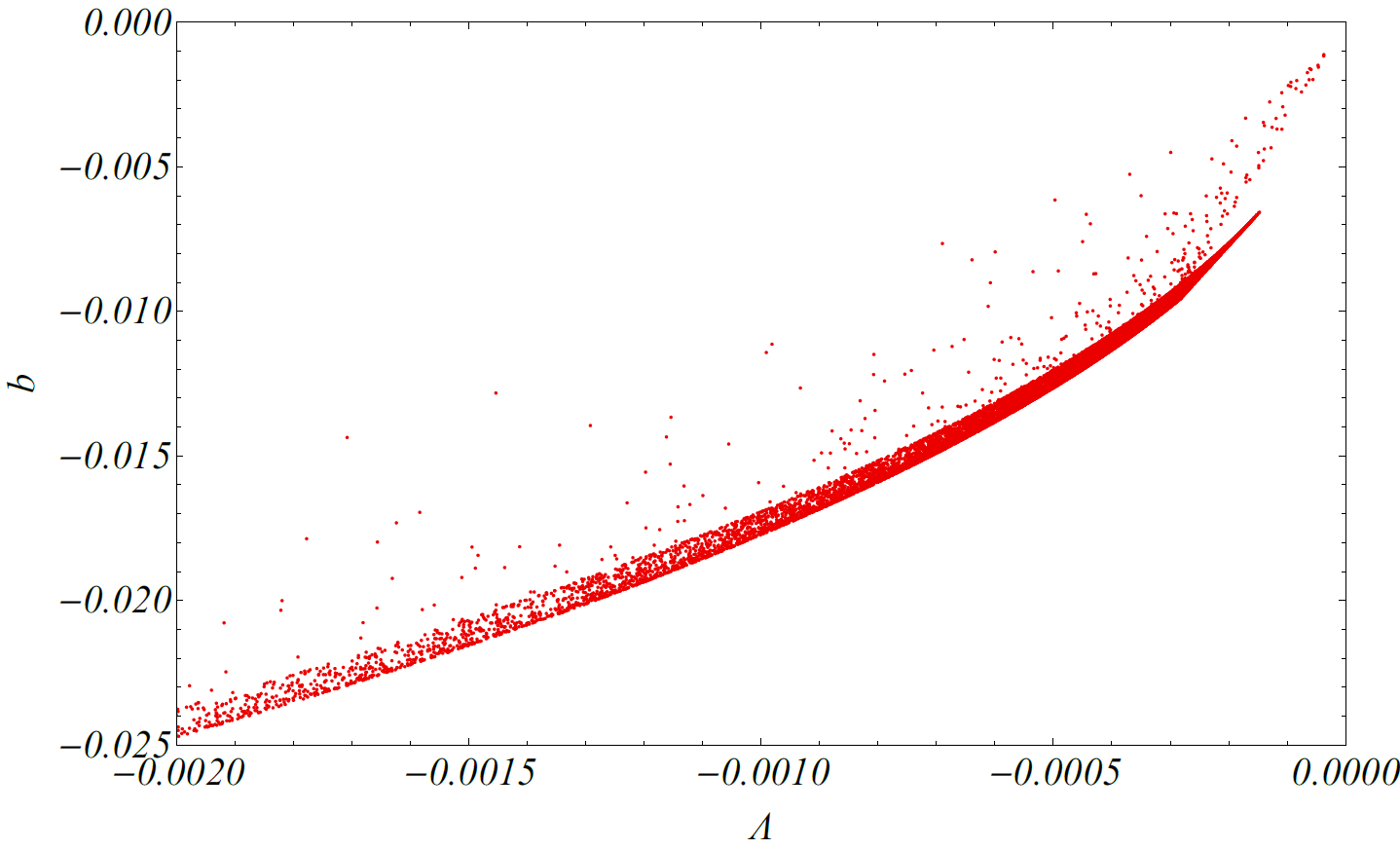} \\
\caption{\small A Monte-Carlo map of the dust matter shell stability parameter region $(b,\Lambda)$ for  $m_{-}=1$, $m_{+}=1.5$. }
\label{fig5}
\end{figure}

Via the same Monte-Carlo process as in the former cases we show a map the stability parameter region in the $(b,\Lambda)$ space (Fig. \ref{fig5}). As is evident in Fig. \ref{fig4} and \ref{fig5} constructed using a Monte-Carlo simulation of the stability range indicated by eqs. \eqref{system_cond}, in both the $w=0$ as well as the $w=1$ case, in order to have the stability conditions satisfied we must have $\Lambda < 0$ and $b < 0$.

In contrast to the potential of eq. (\ref{eq_stiff}) which corresponds to a stiff matter shell, the potential given by eq. (\ref{eq_ordin}) for a pressureless dust shell does not incorporate any high order repulsive terms, \eg $R^4$. This allows for a higher influence of the anti-deSitter term $\sim \Lambda R^2$  ($\Lambda<0$), which is crucial for the implementation of stability at larger R. Therefore, the range of the $(b,\Lambda)$ parameters for which we have stable solutions for the pressureless dust matter case appears to be significantly wider than the corresponding one for the case of stiff matter shell.

\section{Conclusion - Outlook}\label{sectionIV}

We have demonstrated the existence of static, stable spherically symmetric  thin fluid shells in a Schwarzschild-Rindler-anti-de Sitter (SRAdS) metric. We have found analytically the conditions for stability and the corresponding range of values of metric parameters that admit stable fluid shells for different forms of fluid equation of state. These structures have similarities with the well known gravastar shell structures \cite{Mazur:2004fk,Visser:2003ge,MartinMoruno:2011rm,Uchikata:2015yma,Broderick:2007ek}.  In our shell structures the interior de Sitter term of the gravastars  is replaced by a combination of Rindler-anti-de Sitter terms present in a continuous form (same values both in the interior and in the exterior of the shell) allowing for the existence of a minimum of the stability effective potential.

Interesting extensions of this analysis include the following:
\begin{itemize}
\item
The investigation of alternative forms of  metrics that may admit stable shell solutions. For example an interesting alternative simple metric would be one with a Rindler term inside the shell and a Schwarzschild term outside. Such a metric would be free of singularities  and would differ from a gravastar in the replacement of the de Sitter interior by a Rindler interior. Other types of metrics could accept multiple concentric shell structures if the corresponding stability potential has multiple minima at different radii R.
\item
The investigation of observational effects of such shell structures. Since the radius of the shell is always considered in the region outside the event horizon of the black hole,  lensing can be considered in a straightforward manner by studying lightlike geodesics in the SRAdS spacetime along the lines of Refs \cite{Lim:2016lqv, Cutajar:2014gfa} where the lensing of similar metrics is considered. For example signatures of such SRAdS shell structures in typical lensing patterns could be identified and compared to observed lensing patterns around black holes \citep{Hannuksela:2019kle,Bowman:2004ps,Rahman:2018fgy,Ishak:2007ea,Park:2008ih,Sereno:2007rm}. Signatures of SRAdS shells in such optical images  could be specified and compared with predicted signatures of other similar exotic objects like gravastars \cite{Sakai:2014pga}.
\item
The investigation of non-spherical junctions and shells. An interesting problem would be the study of joining rotating spacetimes in the presence of the cosmological constant.
\item
The consideration of more general fluid shell equations of state. In the case of phantom shells it may be possible to have stable shells in a pure Schwarzschild background due to the tendency of such shells to expand rather than contract (negative tension). This is easily shown using the energy conservation equation \eqref{energcons} with $w<-1$ which leads to a surface density  $\sigma(R)=\sigma_0 R^{-2(w + 1)}$ which increases with $R$. The positive value of the exponent for $w<-1$ indicates that it is energetically favourable for such phantom shell to expand rather than contract leading to a negative tension (pressure) that would tend to stabilize the shell even in a pure Schwarzschild background.
\item
The investigation of the dynamical evolution of the shell in the context of spherical symmetry and beyond. Non-spherical dynamical excitations of the shell could also lead to interesting gravitational wave signatures. 
\end{itemize}

\textbf{Numerical Analysis Files}: The numerical files for the reproduction of the figures can be found in \cite{numcodes}.

\section*{Acknowledgements}

The authors would like to extend a thank you to prof. Demetrios Papadopoulos for the stimulating discussions and ideas he provided towards the completion of this paper. This research is co-financed by Greece and the European Union (European Social Fund - ESF) through
the Operational Programme "Human Resources Development, Education and Lifelong Learning 2014-2020" in the context of the project  "Scalar fields in Curved Spacetimes: Soliton Solutions, Observational Results and Gravitational Waves" (MIS 5047648).

\bibliography{bibliography}

\end{document}